\documentclass[10pt, conference, compsocconf]{IEEEtran}
 
\usepackage[numbers]{natbib}
\usepackage{filecontents}

\usepackage{graphicx}
\usepackage{epsfig}
\usepackage{url}
\usepackage{color}
\usepackage[normalem]{ulem}
\usepackage{amsmath}
\usepackage{ulem}

\usepackage{multirow}
\usepackage{slashbox}

\usepackage{natbib}

\newcommand{\send}{\ensuremath{\rightarrow}}
\newcommand{\rev}[2]{{\color{red}\sout{#1}{#2}}}
\renewcommand{\rev}[2]{#2} % OHE: Comment this line to see the revisions à la Word

\begin{document}

\title{Pretty Private Group Management}

\author{
  \IEEEauthorblockN{Olivier Heen, Erwan Le Merrer, Christoph Neumann, St\'{e}phane Onno}

  \IEEEauthorblockA{Technicolor, Rennes, France
\\first.last@technicolor.com}

}

\maketitle

\begin{abstract}
Group management is a fundamental building block of today's Internet
applications. Mailing lists, chat systems, collaborative document
edition but also online social networks such as Facebook and Twitter
use group management systems. In many cases, group security is required
in the sense that access to data is restricted to group members only. 
Some applications also require privacy by keeping group members 
anonymous and unlinkable.
Group management systems routinely rely on a central authority that 
manages and controls the infrastructure and data of the system.
Personal user data related to groups then becomes de facto accessible 
to the central authority.

In this paper, we propose a completely distributed approach for group
management based on distributed hash tables. As there is no enrollment 
to a central authority, the created groups can be leveraged by various 
applications. Following this paradigm we describe a protocol for such a 
system. We consider security and privacy issues inherently introduced by 
removing the central authority and provide a formal validation of 
security properties of the system using AVISPA. We demonstrate the 
feasibility of this protocol by implementing a prototype running on
top of Vuze's DHT.
\end{abstract}

%\begin{IEEEkeywords}
%\end{IEEEkeywords}

\IEEEpeerreviewmaketitle

\section{Introduction}
\label{sec:intro}

Recent years witnessed a rapid growth of Online Social Networks (OSN) sites.
% Today, services such as Facebook and Twitter count several hundred million users (Techcrunch march 2011).
%\footnote{\url{http://techcrunch.com/2010/06/08/twitter-190-million-users/}. Retrieved 03/11.}
%\footnote{\url{http://www.facebook.com/note.php?note_id=112608062113861}. Retrieved 03/11.}
OSNs allow communication with social acquaintances or with users having similar interests.
% Communication means are quite diverse and e.g. consist in sharing photos, chatting, or writing messages. 
\rev{}{The notion of group, also referred to as social order~\cite{Backstrom:2006:GFL:1150402.1150412} is a fairly natural way to sort interactions in our daily social life \cite{Mislove:2007:MAO:1298306.1298311}.}
%\footnote{\url{http://en.wikipedia.org/wiki/Social_order}. Retrieved 03/11.}.
\rev{OSN systems can be considered as group management systems, as they enable one user to cluster its acquaintances and communicate with many users at a time.}{}
% Recently Facebook explicitly introduced the ``Group'' feature which allows to cluster acquaintances according to common interests.
Many of today's \rev{other}{} Internet applications can be considered as group management systems: \rev{}{OSN but also} mailing lists, chat systems or collaborative document edition systems. \rev{These systems either enable one or several users to communicate with many other users at a time or have an explicit notion of group.}{}

From a security and privacy perspective, it is desirable that group information, such as the exchanged messages but also the group member list, is not accessible to anyone outside the group. However, group management systems often rely on a central authority that manages and controls the system's infrastructure and data. Thus, personal user information used during group communication, such as acquaintances, political views and photos, become accessible to the central authority. We believe that an interesting research direction is to give users back control of this data. 

Other drawbacks of group management systems with centralized authority may be mentioned: \emph{(i)} The scalability of the system depends on the capacity of the central authority to dimension the infrastructure resources according to the load. \emph{(ii)} Sharing or reusing groups from one application to another is difficult, as many group management systems define their own proprietary solution and infrastructure. Yet, a group of users subscribed to a VoD service should be able to anonymously contribute to movie reviews on a partner movie site. \emph{(iii)} Bootstrapping new group communication applications requires deployment of a new dedicated infrastructure and system. Therefore, there is a need for a generic and reusable group management mechanism that could be leveraged by various applications dealing with groups.

While contributions have been made for some of the previous problems regarding privacy concerns \cite{Baden:2009:POS:1594977.1592585,secu_api} or regarding scalability issues \cite{Pujol:2010:LES:1851182.1851227}, none of these approaches is able to resolve all above concerns.
In addition, a central authority has low interest in deploying privacy preserving solutions such as \cite{Baden:2009:POS:1594977.1592585,secu_api}.
Therefore, and similarly to Diaspora \cite{diaspora}, a development project for a distributed OSN (see Section~\ref{sec:related}), we believe that only a distributed system with no central authority is able to resolve all of the above issues. 
In such a system the infrastructure is typically composed of a set of end-user devices, which we call nodes, running a piece of software and providing spare storage and CPU resources to the system. 

With a distributed system as described before, we must assume that some participating nodes have been compromised and are under the control 
of an adversary. No central authority can guarantee that the devices running system nodes are honest. In such a context, we are thus interested in building a distributed group management system with fair security and privacy properties against participating nodes.
Whisper \cite{Schiavoni2011} has similar objectives and specifically focuses on confidential communication within formed groups. Whisper achieves them by combining gossip-based communication protocols and onion routing. In this paper, we focus on a broader set of services that allow building a complete group management system.
%The solution however requires each participant to contribute with a node in the communication system. 
On the security side, Whisper considers a threat model where nodes fully comply with the specified protocol but try to passively eavesdrop member and group information or any other type of message not meant to be read by this particular node.
Whilst we consider the same threat model regarding privacy properties, we extend the model to a Dolev-Yao adversary \cite{Dolev:1981:SPK:891726} regarding security properties and formally validate our security objectives against the latter attacker.

{\bf Contributions:} we propose a completely distributed approach for group management requiring no central authority, based on Distributed Hash Tables (DHTs). 
The proposed system is generic and may be leveraged by various applications that need group management. 

\rev{Even though the system is distributed it}{The system} offers a set of security properties against adversaries that compromised nodes of the system.
This includes confidentiality and integrity of group information against a Dolev-Yao adversary \cite{Dolev:1981:SPK:891726}.
In addition the system enables user anonymity and ensures that users belonging to two different groups stay unlinkable, in particular against nodes of the DHT that comply with the protocol specification but try to passively steal information of other participants or groups. If required by group policy, anonymous communication between group members is also possible.

We provide a formal validation of security properties of the protocol using AVISPA~\cite{springerlink:10.1007/11513988} and discuss to what extent the proposed system meets the privacy objectives.
The proposed system does not try to address attacks on the system availability, such as Byzantine failures.
Instead, we refer to the recent advances in this field \cite{Young:2010:PRC:1845878.1846270}.
We however address the specific security and privacy problems that have been introduced by removing the central authority and by moving to a completely distributed architecture.

Finally, we prototyped our protocol on top of the Vuze DHT. While limitations remain because of the non-optimized Vuze DHT, we show that distributed and private group management is feasible with acceptable performances.

The remainder of this paper is structured as
follows. Section~\ref{sec:distributed} introduces the proposed group
management system with no security. We then present in
Section~\ref{sec:secprot} our security objectives and considered
adversaries, the cryptographic means and security protocol used. In
Section~\ref{sec:seccons} we analyze the security and privacy
properties of the system. We then present in
Section~\ref{sec:prototype} a prototype of our system, and the results
of a $2.5$ day execution. Related work is presented in
Section~\ref{sec:related}. Finally we conclude with
Section~\ref{sec:conclusion}.

% Formalization of an access control model for online social networks.

\section{A distributed group management system}
\label{sec:distributed}
\subsection{Motivating example}
\label{ssec:example}
We propose collaborative document edition as an example that motivates
the needs of a social-based and secure application.  
Collaborative document edition is an example where users
want visible actions to be restricted to a trusted group of people;
however, the decentralized location of the edited document is also
of interest for privacy, as opposed to storage at a single service
provider.

Our group management system allows the creation of a group
dedicated to specific documents.
Then it allows the control of who should be able to join the group of 
people editing the document.
The group(s) administrator(s) grants access to a user. Depending on the document policy editions could be performed anonymously. It is also possible to hide which pseudonym made the modifications.
Modifications are saved and
accessible on nodes participating to the group management system,
in a distributed fashion, thus potentially allowing a large number of
users to edit the same document. Users that are banned from the group of
editors, users that decided to leave, or any other non group member, are not
able to perform editions, nor to read the document.

%We now describe the core of our group management system, and precise
%how such an example application can be implemented. We restrict to
%asynchronous edition for sake of clarity.

\subsection{System schematic}

\begin{figure}[t!]
\begin{center}
\includegraphics[width=0.9\linewidth]{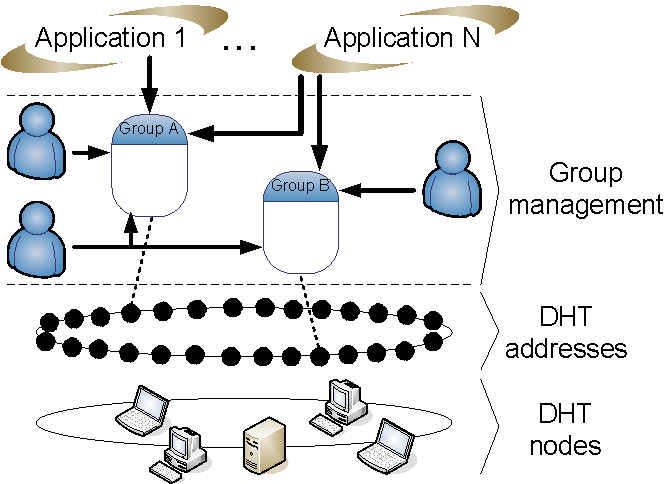}
\end{center}
\caption{High level view of our group management system:
    multiple applications are accessing groups managed by
    users. Groups are mapped to random DHT locations, and are hosted
    on commodity hardware.}
\label{gman}
\end{figure}

A distributed hash table is a well known tool when it comes to
decentralized and scalable 1 to 1 communication. It exports a basic
interface providing PUT and GET operations, allowing to map
$<key,value>$ pairs to nodes participating in the system. This is done
by hashing an object's content, in order to obtain a random address on
the DHT's address space (typically of size $160$ bits, as
\textit{e.g.} in the Vuze's DHT). Nodes are themselves responsible for
a subset of this space, based on their position in the DHT (depending
on their ID in the same address space).

\rev{}{In this paper, we assume a Byzantine Fault Tolerant DHT for building 
our system.
Indeed,} recent advances make DHTs tolerant to Byzantine adversaries, while 
conserving logarithmic cost of operation in expectation~\cite{Young:2010:PRC:1845878.1846270}.
Basic storage systems on top of DHTs~\cite{Rowstron:2001:SMC:502034.502053}
can also implement Byzantine fault tolerant replication in relatively
stable environments, using mechanisms providing eventual
consistency~\cite{Singh:2009:ZEC:1558977.1558989}.
Nevertheless,
handling both dynamicity and BFT nodes is still under
research~\cite{Bao:2009:QOC:1586637.1587153}.

\rev{In this paper, we assume a BFT DHT, on top of which we build our system; we will
discuss in Section~\ref{priv:disc} that if in the worst case some DHT
nodes are compromised, our system ensures that information stored on
them is unreadable.}{}

We design our group management system around a \textit{group
  structure}.  This abstraction, composed of a \textit{root}, a member
\textit{list}, a \textit{wall} and an \textit{inbox}, allows the
representation of complex objects and interactions.  A user that joins
the system can create \textit{principals}. We consider a principal as
an instance of a group structure, the list being for instance in the
context of OSNs filled with friends, and the wall by messages for this
given principal; the inbox receiving system or user messages. Users
can also create and manage groups, which contain other users, groups
and objects. Belonging to a group, thus being a \textit{member}, means
being able to access objects of that group and interact with other
group members, depending on the defined groups privacy and security
policy. Users create principals to join with groups, possibly one new
principal for each joined group.  Group \textit{creators} can create
and destroy groups and define the join policy and the visibility
policy of the group, as illustrated in Table~\ref{visibility}.
\textit{Administrators} handle join and leave requests.  The great
flexibility regarding the join policy and the visibility enables very
different types of applications to run on top of the system.  The roles
\textit{creator}, \textit{administrator} and \textit{member} depend on
the knowledge of cryptographic keys described in
Section~\ref{sec:cryptomeans}.

%% \begin{table}
%% \small
%%     \begin{tabular}{|c|c|c|c|}
%%         \hline
%%  Join Policy  &&&\\
%%  & Anyone & Members & Administrator \\
%%  List or Wall Visibility &&& \\ \hline
%% \multirow{3}{*}{Anyone may join} & open Facebook group & chat room & survey, newsletter \\ 
%% & petition & Facebook group & RSS feed \\ 
%% \hline
%% \multirow{3}{*}{Selected join} & program committee & picture sharing & subscription to VoD\\
%% &  & outlook invite & user of a transport system \\
%%  \hline
%%     \end{tabular}
%% \vspace{0.2cm}
%% \caption{Levels of group visibility, and examples of application.}
%% \label{visibility}
%% \end{table}

\begin{table*}
\center
\footnotesize
    \begin{tabular}{|c||c|c|c|}
        \hline
 
 \backslashbox{Join Policy}{List Visibility} & Anyone & Members & Administrator \\
  \hline \hline
\multirow{3}{*}{Anyone may join} & open Facebook group & chat room & survey, newsletter \\ 
& petition & secret Facebook group & RSS feed \\ 
\hline
\multirow{3}{*}{Selected join} & program committee & picture sharing & subscription to VoD\\
&  & outlook invite & transport system \\
 \hline
    \end{tabular}
\vspace{0.2cm}
\caption{Levels of group join policy and list visibility, and examples of application.}
\label{visibility}
\end{table*}

On an implementation level, the \textit{root} is the entry point of a
group structure; it is a file containing metadata about group's
attributes, and pointers to \textit{list} and \textit{wall}. The \textit{list}
references principals and groups that are members of the current
group. The \textit{inbox} is a list of messages, typically join request from principals. 
Finally, the \textit{wall} is to be seen as a space containing
raw data as objects (if their size is small) or references to objects,
and system messages. On the distributed side, each
element of a group structure, as well as nodes' inbox, is hosted on a
random DHT node (and then replicated on node's neighbors for
reliability~\cite{Rowstron:2001:SMC:502034.502053}). Figure \ref{gman}
presents a high level view of the system, with applications leveraging
groups of users.

This system is made available to programmers through an API, providing
basic operations to create, manage, join, leave, list members of a
group, or to send a message to a given principal or to the whole group
for instance.

\subsection{Benefits and properties}

As opposed to the design of an access management for a particular
application, the purpose of abstracting the notion of group as a
general system-core entity is to provide means of genericity,
reusability and applicability. Our group management system is
\textit{generic} in the sense that the operations it provides are
general yet powerful enough to be leveraged by multiple social-based
applications. When a group is formed in our system, it can be accessed
by different applications, providing \textit{reusability}. In other
words, a group instance can be used by multiple applications at
runtime, without the need for those applications to collaborate or to
be aware of their respective existence.  Finally,
\textit{applicability} comes from the fact that our system can be run
on commodity hardware, avoiding the need for investment in a server
farm or rental of a specific cloud service. As it is by nature
distributed, it can be hosted on user machines, along with the
application that is using it.  
Finally, relying on a DHT
  and distributing responsibilities to random nodes allows our system to be scalable in the number of users and in the number of groups. Scalability in the number of users per group can however be an issue for very large groups, as nodes hosting structures may be contacted frequently. However, studies \cite{Mislove:2007:MAO:1298306.1298311} reveal that group sizes are following a power-law distribution, with a vast majority of groups containing only few members.
For very large groups, we do not claim to provide better scalability than traditional distributed applications as for instance \textit{publish/subscribe} systems, also relying on master nodes in DHTs~\cite{10385790}.

\section{Security protocol}
\label{sec:secprot}
\subsection{Security and privacy objectives}
\label{sec:secobj}
% We do not consider this type of attack in this work.

% Centralized group management systems provide users credentials and administrates groups from a single access point (e.g. www.facebook.com). The whole elements and means that compose such system are under control of a central entity. There are so-called generic attacks on such system such as (TODO REF) authent (TODO REF) farming. 

% A distributed group management system spreads the system over independent nodes of a distributed hash table. The control is distributed among the group members that own the groups. The physical nodes that provide part of the system are hosted on end-user computers. The computer, thus the node might be on hands of an attacker who can forge confidentiality, integrity or availability attacks. Similarly, privacy properties (REF TODO PFITZMANN) such as anonymity and unlinkability might be vulnerable. Therefore, a distributed management system must consider these specific attacks from these malicious nodes. 

In this work we focus on the attacks that become possible because of the distributed nature of our group management system: 
some nodes providing storage and CPU resources could be under the control of an adversary.
%We consider the following cases: i) One node is under the control of the attacker ii) $f$ nodes are under the control of the attacker and iii) all nodes of the DHT are under the control of the attacker.
Attacks leading to user profiling and De-anonymization \cite{Balduzzi_abusingsocial,10.1109/SP.2010.21} have been demonstrated against deployed OSNs. These attacks rely on publicly available group member lists. Our system does not claim additional resistance to such attacks for groups with publicly available member lists.
% They do not depend from the group management system being distributed or under the control of a central authority.
% Our system does not claim additionnal resistance to such attacks.

Our security objectives are relative to the Dolev-Yao (DY) adversary \cite{Dolev:1981:SPK:891726}. 
The DY adversary fully controls the network and some nodes but can not reverse any cryptographic operation. Our security objectives are:
\begin{itemize}
\item Ensure the confidentiality of private and secret keys, see table \ref{tab:structures}.
\item Ensure the access to public keys according to group policy. 
\item Ensure access control to group information (wall, list, membership) according to group policy.
\item Ensure the integrity of messages sent by participants.
\item Ensure the security of the capture and update mechanism (see \ref{sec:cryptomeans}).
\end{itemize}
Regarding availability, the present work does not address Byzantine failures. We refer to the recent advances in this field \cite{Young:2010:PRC:1845878.1846270}. 
We however discuss the risk of an adversary flooding the DHT or squatting addresses\footnote{Note that the DY adversary is not used here as he succeeds in blocking {\it any} protocol.}.

% In this model the attacker is able to listen, intercept, and modify any message that flows through a  node he controls. The attacker is only limited by cryptographic method that he is not able to break.
% The security objectives we analyze are {\em Confidentiality}, {\em Integrity} and {\em Availability}:
%\begin{itemize}
%\item {\em Confidentiality}: Only users that explicitly received corresponding key material shall be able to read information exchanged and stored within a group. This information includes the wall data, member lists and messages of inboxes. Even nodes storing information for a given group shall not be able to access this information in clear-text. 
%\item  {\em Integrity} : An attacker shall not be able to alter an existing information message originated by a user or node.
%wall data, member lists and messages of inboxes shall only be modifiable by users that have the according right. Any unsolicited changes should be detectable. In addition it should not be possible for an attacker to claim ownership of a resource that he does not possess.
%% and thus delete this resource. 
%Note that, we consider that a Byzantine fault tolerance replication ensures the overall system replication consistency.
%% , which thus detects individual nodes modifying locally only the data they store.
%\end{itemize}

The privacy objectives listed below are relative to an adversary that adheres to the protocol but uses information from controlled nodes and observed messages to gain additional information.
Following the terminology of Pfitzman et al. \cite{springerlink:10.1007/3-540-44702-4}, the privacy objectives are:
\begin{itemize}
\item{Members anonymity}: the adversary can not retrieve the identity of a group member.
\item{Senders anonymity}: the adversary can not retrieve the sender of a private message.
\item{Unlinkability with IP address}: the adversary can not associate group members with IP addresses, and thus use the IP address as an identifier.
\item{Members unlinkability}: the adversary shall not link two identities in the system. In particular he shall not infer that a principal is member of two different groups.
% \item{\em Structure unlinkability}: the adversary shall not be able to link the structure he stores to other structures of the same group.
\end{itemize}
Our system does not provide {\em unobservability}. An adversary may infer group information like estimating cardinality or the frequency of actions. However we discuss how a sufficiently large DHT together with the PUT/GET mechanism could complicate the observability.
Note that there exist obvious limitations to the above privacy objectives: trivial anonymity sets, \rev{users}{principals explicitly} revealing identities, etc. We do not address those issues in the present work.

\subsection{Cryptographic means}
\label{sec:cryptomeans}
A set of mechanisms allows achieving the security and privacy
objectives described in the previous section.  First, the usage of
{\em cryptographically generated address} (CGA)
\cite{Montenegro:2004:CIC:984334.984338,springerlink:CGA} on the DHT
ensures that only the owner of a given public private key pair is able
to control the calculated address.  Second, each node of the DHT
verifies the signatures of data stored at an address that the node
currently hosts. The nodes deny updates if the data is not signed with
the private key used by the CGA mechanism. We call this mechanism the
secure address capture and a secure update mechanism.  Finally, users
of the group communication system do not need to run individual
nodes. Instead, the users use PUT and GET operations to write and
retrieve data at specific addresses in the DHT.  Thus, users never
communicate directly with each other, which allows keeping them
anonymous. This concept is similar to the usage of post-office boxes
in the postal system.

All data \rev{Each type of}{} structures \rev{{\em root}, {\em list}, 
{\em wall} and {\em inbox} introduced in Section~\ref{sec:distributed}
uses its own}{have a} set of cryptographic keys.
Public-private key pairs ensure the structure's integrity and are used 
to distribute write permissions to the users.
Symmetric keys ensure the structure's confidentiality and are used to
distribute read permissions to the users.
The {\em cryptographically generated address} (CGA) is calculated using 
a hash function, noted $h()$ hereafter, on the structure's public key.
Cryptographically generated addresses ensure that only the owner of a
given public private key pair is able to control the calculated address.

% The advantage of using CGAs are twofold: i) It reduces the risk of attackers squatting unused addresses in the DHT and ii) it allows users and nodes to systematically verify the correct location of a structure. The latter reduces the risk of luring users to a fake address that an attacker gained control of.

\rev{At start, all the addresses of the DHT are free and anyone
can capture a free address by providing corresponding key
material. Later on, only the user that captured an address, can
perform operations on this address. This mechanism is implemented and
enforced at node level by systematically verifying the signatures of
structures and denying updates if the structures are not correctly
signed. In addition a structure's integrity is systematically verified
by users of the system.
}{All structures of type {\small {\tt root}}, {\small {\tt list}} and {\small {\tt wall}} are self-signed using the structure's public and private key pair $K,K^{-1}$.
In order to allow verification of the signatures by the storing nodes themselves and by anyone retrieving the structure, the public key $K$ is also stored in clear-text at the structure's storage address $h(K)$.
Each self-signed structure has a counter $c$ that is incremented at each update to prevent replay attacks.

The address capture is successful when the storing node verifies that the address is empty and $c=0$.
The update is successful when the storing node verifies that the address is not empty and the signing key is unchanged and the counter is correctly incremented\footnote{The increment is modulo the width of the counter, a long integer in our implementation.}.
}

The inbox structure is not self signed as a whole and not subject to \rev{}{the capture and update} mechanism.
This allows anyone writing into the inbox.
However each message is self-signed using the sender's keys.
In order to preserve the sender's anonymity against the storing node, the sender's public key is encrypted within the sent message using the receiver's public key (see Section~\ref{sec:protocol}).

\begin{table*}
\begin{center}
\begin{tabular}{l|c|c|c|c|c}
Structure & Storage address& Clear-text & Signing key & Encryption key & Counter\\
 \hline
Root& $h(K_{r})$ & $K_{r}$&$K^{-1}_{r}$ & None & $c_r$\\
List& $h(K_{l})$ & $K_{l}$&$K^{-1}_{l}$ & $S_l$ & $c_l$\\
Wall& $h(K_{w})$ &$K_{w}$ &$K^{-1}_{w}$ & $S_w$ & $c_w$\\
Inbox& $h(K_{i})$  &None & Sender's private key & $K_{i}$ & none\\
\end{tabular}
\vspace{0.2cm}
\caption{Group structures cryptographic means.}
\label{tab:structures}
\end{center}
\end{table*}

Table~\ref{tab:structures} gives an overview of the different keys and
addresses used by the system.  The root structure is not
encrypted. Thus, any user knowing the public key $K_{r}$ or the
address $h(K_{r})$ is able to retrieve the root structure. However,
the root structure's integrity and write protection is ensured by the
public private key pair $K_{r},K^{-1}_{r}$. $K_{r}$ is stored in
clear-text at the address $h(K_{r})$, which allows nodes and users to
verify the integrity and correct location of the structure.  The
member list is encrypted with a key $S_{l}$ and signed by the key
$K^{-1}_{l}$. Any user having the key $K^{-1}_{l}$ and $S_{l}$ can
update the list. Any user having the key $S_{l}$ can read the member
list.  Similarly to the root structure, $K_{l}$ is stored in
clear-text at the address $h(K_{l})$. The wall is encrypted with key
$S_{w}$ and signed by the key $K^{-1}_{w}$.  Anyone knowing $S_{w}$
can read the data on the wall. Anyone having $K^{-1}_{w}$ and $S_{w}$
can write on the wall. $K_{w}$ is stored in clear-text at the address
$h(K_{w})$.  Finally, the inbox is not protected in integrity. However
each stored message in the inbox is encrypted with the public key
$K_{i}$ of the inbox. In addition, the sender of a message also signs
the message with its private key.

Table~\ref{tab:roles} summarizes keys required for each the roles
introduced in Section~\ref{sec:distributed}. Members receive keys for
a given group according to the group policy.

\subsection{Protocols}
\label{sec:protocol}
We describe the main protocols of our system using the common Alice \& Bob notation.
In this notation the statement "$x$ sends the message $m$ to $y$" is denoted $x \send y : m$.
To denote a message $m$ encrypted by a key $K$ we note $\{m\}_{K}$. To denote a message $m$ signed by a key $K^{-1}$, we use the compact form $\{m\}_{K^{-1}}$ instead of the longer $m.\{h(m)\}_{K^{-1}}$. This hides the construction for existential unforgeability.
This also stresses that the signature does not protect confidentiality.
We note $ x \send dht(a) : m $ when $x$ performs the operation $PUT(a,m)$ over the DHT.
We note $ dht(a) \send x : m $ when $x$ performs the operation $m=GET(a)$ from the DHT.
We denote a list as $[,]$. Finally we use $a.b$ for the concatenation of $a$ and $b$.

Using this notation the general form of the capture mechanism is: 

\begin{center}
$x \send dht(h(K)) : \{\tt type.c.K.payload\}_{K^{-1}}$.
\end{center}

The strings {\small{\tt root}}, {\small{\tt list}}, {\small{\tt once}}, {\small{\tt wall}}, {\small{\tt helo}}, {\small{\tt name}} and {\small{\tt join}} denote message types.

\subsubsection{Creating a group}
Creating a group mainly consists of capturing the DHT addresses for components (root, list, wall) and publishing the group name in a directory.
\rev{}{First the group creator generates a set of keys: $(K_r^{-1},K_r), (K_i^{-1},K_i), (K_l^{-1},K_l)$, $(K_w^{-1},K_w), S_l, S_w$ and plays the group creation protocol as follows:}
\begin{align}
\small
g \send dht(h(K_r))	&: \{\{{\tt root}.c_r.K_r.K_i\}_{K_i^{-1}}\}_{K_r^{-1}}	\\
g \send dht(h(K_l))	&: \{{\tt list}.c_l.\{[ ]\}_{S_l}.K_l\}_{K_l^{-1}}	\\
g \send dht(h(K_w))	&: \{{\tt wall}.c_w.\{[ ]\}_{S_w}.K_w\}_{K_w^{-1}}	\\
g \send dht(h(K_i.0))	&: \{{\tt once}.\{c_l\}_{K_l^{-1}}\}_{K_l}	\\
g \send directory	&: \{{\tt name}.K_r\}_{K_r^{-1}}			%\\
\end{align}
\setcounter{equation}{0}

Messages (1) (2) (3) set-up the data structure for the group
Message (4) stores a signed and encrypted version of the list counter $c_l$ at the address $h(K_i.0)$.
This counter is used as an anti-replay protection for join requests as shown in the join protocol below.
We choose the address $h(K_i.0)$ because it depends from $K_i$ and because it does not override the inbox $h(K_i)$.
Message (5) publishes the group name in a directory.

\subsubsection{Creating a principal}
As indicated in Section~\ref{sec:distributed}, a principal is a group possibly with no wall and no member list.
A principal willing to remain anonymous will not publish the public key of her inbox and not publish any information into a directory.
First the user generates a set of keys: $K_p^{-1},K_p$.
\begin{align*}
\small
p \send dht(h(K_p))	&: \{{\tt root}.K_p\}_{K_p^{-1}}			%\\
\end{align*}
\setcounter{equation}{0}

\subsubsection{Joining a group}
This is the most important operation in our group management system.
\rev{}{First the principal generates key pair $K_j^{-1},K_j$.
$h(K_j)$ is an inbox for receiving messages from the administrator.
In our example, $h(K_j)$ is also used for receiving messages from other group members.
Note that other applications may use two different inboxes here.}

Then the operation has three main stages.
%A principal willing to join retrieves the group information from a directory and from the root address.
%Later on, any entity knowing the group list private key $K_l^{-1}$ may then process the request.
%After a delay, the principal retrieves the membership information at the reply address $h(K_j)$.

\begin{itemize}
\item The principal puts a join request.
\begin{align}
\small
directory \send p	&:\{{\tt name}.K_r\}_{K_r^{-1}}			\\
dht(h(K_r)) \send p	&:\{\{{\tt root}.c_r.K_r.K_i\}_{K_i^{-1}}\}_{K_r^{-1}}	\\
dht(h(K_i.0)) \send p	&:\{{\tt once}.\{c_l\}_{K_l^{-1}}\}_{K_l}		\\
p \send dht(h(K_i))	&:\{\{\{{\tt join}.K_p.K_j.\{{\tt once}. \nonumber\\
& \{c_l\}_{K_l^{-1}}\}_{K_l}\}_{K_p^{-1}}\}_{K_j^{-1}}\}_{K_i}					%\\
\end{align}
% The use of a counter $c$ in message (3) is detailed below.

\item An administrator $a$ gets and processes the join request.
\begin{align}
\small
dht(h(K_i)) \send a &: \{\{\{{\tt join}.K_p.K_j.\{{\tt once}.\nonumber\\
&\{c_l\}_{K_l^{-1}}\}_{K_l}\}_{K_p^{-1}}\}_{K_j^{-1}}\}_{K_i}					\\
dht(h(K_l)) \send a &: \{{\tt list}.c_l.\{[X]\}_{S_l}.K_l\}_{K_l^{-1}}	\\
a\send dht(h(K_i.0))&: \{{\tt once}.\{c_l+1\}_{K_l^{-1}}\}_{K_l}	\\
a \send dht(h(K_l)) &: \{{\tt list}.c_l+1.\{[X,(K_p,K_j)]\}_{S_l}.\nonumber\\
&K_l\}_{K_l^{-1}}.K_l \\
a \send dht(h(K_j)) &:\{\{{\tt helo}.[keys]\}_{K_i^{-1}}\}_{K_j}		%\\
\end{align}

\begin{table*}[t!]
\begin{center}
\begin{tabular}{l|c}
Role & Required keys \\
 \hline
Creator&  $K^{-1}_{r}$ \\
Administrator& $K_{i}$, $K^{-1}_{i}$, $K^{-1}_{l}$, $K^{-1}_{w}$, $S_l$, $S_w$ \\
Member& $S_l$, $S_w$, $K^{-1}_{w}$ depending on group policy\\
\end{tabular}
\vspace{0.2cm}
\caption{Group roles and associated keys.}
\label{tab:roles}
\end{center}
\end{table*}

\item The principal retrieves the group information.
\begin{align}
dht(h(K_j)) \send p	&:\{\{{\tt helo}.[keys]\}_{K_i^{-1}}\}_{K_j}		%\\
\end{align}
%The possibles values of $[keys]$ in message (10) are given bellow.
\end{itemize}
\setcounter{equation}{0}

The counter $c_l$ is used in messages (3) and (7) to prevent replay attacks.
It is signed and encrypted by administrators and used as a ticket in a join request.
Upon processing the join request an administrator checks that the counter value corresponds to the counter value of the list (in fact, strict equality is not required, it is sufficient that the counter is greater or equal than the current list counter).
% This mechanism is important for preventing replay of the join request (see Section~\ref{sec:justification}).

In message (6), $[X]$ is the current list of members.
The administrator sends message (8) for adding $K_p$ in the list and update the counter accordingly.

In message (9) and (10) the value of $[keys]$ depends on the group policy.
The minimum is $[]$ for a totally private group, typically for subscriptions to catalogs (see Table~\ref{visibility}).
The maximum is the full list of keys for a totally open group.
For the example of Section~\ref{ssec:example} the key list is $[S_w,K_w^{-1}]$. 
This lets anyone read/write the wall and keeps the user list private.

\subsubsection{Taking actions in the group}
After joining a group, a principal may enjoy group activities.
In our example a principal will anonymously contribute to the shared document.
We show the protocol exchange to do so in a minimalist model of a shared document (that just allows read and replace).
\begin{align*}
\small
dht(h(K_w)) \send p	&:\{{\tt wall}.c_w.\{[old]\}_{S_w}.K_w\}_{K_w^{-1}}		\\
p \send dht(h(K_w))	&:\{{\tt wall}.c_w+1.\{[new]\}_{S_w}.K_w\}_{K_w^{-1}}		%\\
\end{align*}

\subsubsection{Public communications}
Anyone knowing $h(K_i)$ may write a message in the corresponding inbox.
\begin{align*}
\small
p \send dht(h(K_i))	&:{\tt mess}.message				%\\
\end{align*}
Such communication can not be avoided in environments with no central authority.
Optimistically this is an opportunity to contact principals that publish their address $h(K_i)$.
Pessimistically \rev{}{this} is spam.
Note that in our motivating example $h(K_i)$ is never disclosed nor {\it a fortiori} $K_i$, thus limiting the risk of spam.

\subsubsection{Private communications}
Our system allows private communications within users of a group.
According to the group policy, members may learn the inbox key $K_i$ of other members directly from the member list, or through trusted external channels like direct communication between people.
A private communication is thus systematically encrypted using the key $K_i$, which makes it fundamentally different from an open communication.
\begin{align*}
\small
p \send dht(h(K_i))	&:\{\{{\tt mess}.message\}_{K_p^{-1}}\}_{K_i}		%\\
\end{align*}
In addition, a known mechanism can be used for hiding the IP address of the sender $p$ to several adversary nodes. We provide here an example of such mechanism, directly adapted from Crowds \cite{Reiter:1999:AWT:293411.293778}.
$\alpha$ and $\beta$ are random addresses, messages (2) is sent with probability $p_f>1/2$ message (3) is sent otherwise and terminates the protocol.
\setcounter{equation}{0}
\begin{align}
\small
p \send dht(\alpha)	&:h(K_i).\{\{{\tt mess}.message\}_{K_p^{-1}}\}_{K_i}	\\
dht(\alpha)\send dht(\beta)&:h(K_i).\{\{{\tt mess}.message\}_{K_p^{-1}}\}_{K_i}	\\
dht(\alpha)\send h(K_i)	&:\{\{{\tt mess}.message\}_{K_p^{-1}}\}_{K_i}		%\\
% dht(\alpha){ 
  % \begin{array}{l l}
	% \send dht(\beta)&:h(K_i).\{\{{\tt mess}.message\}_{K_p^{-1}}\}_{K_i} & \quad \text{if $n$ is even}\\
    % \send h(K_i)	&:\{\{{\tt mess}.message\}_{K_p^{-1}}\}_{K_i} & \quad \text{if $n$ is odd}\\
  % \end{array} 
\end{align}
\setcounter{equation}{0}

\subsubsection{Key renewal}
An administrator can decide to renew keys such as $S_w$. The reason for a key renewal may be the banishment of a group member. The administrator sends the new keys to the inbox of each group member except the banned member. This is possible because the administrator knows the list of members and their inboxes $K_j$. Other more complex cases have to be considered such as cases where the administrator cannot directly address the group members (e.g. members haven't revealed their $K_j$). Further work will investigate the key renewal mechanisms in such cases.

% LocalWords:  abusingsocial anonymization OSNs beeing additionnal Dolev Yao

\section{Security and Privacy Considerations}
\label{sec:seccons}
% This section presents the different attacker models and security and privacy objectives that we consider. We also formally validate our protocol regarding security objectives and discuss our privacy properties.

As such the proposed system is a system without authentication. Any user may create its groups and principals and associated keys. No central authority and in particular no public key infrastructure (PKI) is required, which makes the system scalable. Another advantage is that anyone can participate by creating it's principals and groups.
Fundamentally, this is not different from services such as Wikipedia where anyone may sign in and contribute without authenticating, or webmail services such as Hotmail or Yahoo Mail, where anyone can create as many accounts as he wants without authenticating. The disadvantages of such systems are that it is possible to squat certain addresses or to flood the address space.
As discussed in Section~\ref{sec:distributed} the address space of a DHT is typically $2^{160}$. We consider that an exhaustive flooding of the entire address space of the system is prohibitive. This cost of flooding may also be increased using computational puzzles, in a fully distributed and scalable way~\cite{Borisov:2006:CPS:1157740.1158254}.  
In addition the usage of CGAs and the address capture mechanism (see Section~\ref{sec:cryptomeans}) reduces the risk for an adversary to squat a particular address in the DHT.
% and ii) it allows users and nodes to systematically verify the correct location of a structure. The latter reduces the risk of luring users to a fake address that an attacker gained control of.
Finally, distributed systems with no authentication are subject to Sybil attacks \cite{Cheng:2005:SRM:1080192.1080202} and solutions such SybilGuard~\cite{Yu:2006:SDA:1159913.1159945} deal with this attack. Sybil attacks are out of scope of the present work.

It is also worth noting that the application plugged on top of our group management system may implement its own authentication mechanism, based e.g. on PKIs, mail-address checking or captchas, thus controlling the users accessing the underlying system.

% \section{Justification}
\subsection{Security analysis}
\label{sec:justification}

We first verify the capture and update mechanism.
More precisely we verify that a DY adversary is not able to update a captured address, unless she captured it herself.
Within the AVISPA framework \cite{springerlink:10.1007/11513988}, we provide a formal specification of the capture mechanism as well as security goal for the weak authentication of the entity that captures the address\footnote{The AVISPA code is included in appendix.}.

The simulation shows that the unpredictability of the captured addresses is critical.
If the DY adversary does not know a public key prior to the capture of the corresponding address, no attack is found.
Otherwise, for instance if the adversary knows a key $K_l$, the attack bellow exists (messages (1) to (5)).
\setcounter{equation}{0}
The adversary $i$ turns a predicted address into an inbox, so that it accepts any further message without verification:
\begin{align}
i \send dht(h(K_l))	&: {\tt mess}.message				%\\
\end{align}
The group captures (2) and uses (3) (4) the predicted address without noticing any difference:
\begin{align}
g \send dht(h(K_l))	&: \{{\tt list}.0.\{[ ]\}_{S_l}.K_l\}_{K_l^{-1}}	\\
g \send dht(h(K_l))	&: \{{\tt list}.1.\{[p_1]\}_{S_l}.K_l\}_{K_l^{-1}}	\\
g \send dht(h(K_l))	&: \{{\tt list}.2.\{[p_1,p_2]\}_{S_l}.K_l\}_{K_l^{-1}}	%%\\
\end{align}
The adversary $i(g)$ pretending to be $g$ replays one former message that will be accepted without signature and without increment verification.
Here, the effect is the unauthorized removal of the principal $p_2$ from a group:
\begin{align}
i(g)\send dht(h(K_l))	&: \{{\tt list}.1.\{[p_1]\}_{S_l}.K_l\}_{K_l^{-1}}	%\\
\end{align}

We also model the protocol for creating a group, then creating a principal, and then joining the group (see the second portion of code in the appendix).
We assume a secure channel between the group creator and the future administrators.
This channel is used for transferring the keys $(K_l^{-1},K_l), S_k, (K_i^{-1},K_i)$.
The assumption is reasonable when an administrator is the group creator itself, or when a secret is shared (which we have modeled in the simulation).
It is also possible that a creator and some administrator belong to a same private group.

We systematically verified the secrecy of the private keys and the
symmetric keys against two different kinds of DY adversaries.  They
both control the messages send over the network.  The first controls
all the addresses from the DHT.  The second controls all addresses
except those involved in the management of the group and the
principal; note that this adversary still controls all inbox addresses
as well as addresses of type {\tt once}.

For private groups, as the group in our toy example Section
\ref{ssec:example}, we obtain the secrecy of the group key $K_l$
against the two types of adversaries.  We obtain the secrecy of the
keys of the principal $K_i$ and $K_p$ against the second type of
adversary.

%We did not formally verifying additional security properties using the same portion of code, %although weak authentication properties should be verifiable.
%As a further work, we envision a complementary model within another framework like Scyther \cite{Scyther}.

% {\CNE Do not forget: Detection of nodes changing stored information thanks to BFT.}
\subsection{Privacy discussion}
\label{priv:disc}

We now discuss to which extent our protocol meets the privacy objectives discussed in Section~\ref{sec:secobj} with an adversary that adheres to the protocol but uses information from controlled nodes and observed messages to gain additional information.

Some of the privacy objectives are achieved thanks to the confidentiality of information. {\em Member anonymity} would be broken if the adversary retrieved the public keys $K_i$ or $K_p$ from group member lists, walls or inboxes. 
% Both anonymity properties are achieved thanks to confidentiality. 
% {\em Member anonymity} would be broken if an administrator's inbox, a group member list or the wall were readable for an adversary not member of the group. 
However, these structures are encrypted and only accessible to the group members or group administrators. Thus, a single node storing a member list, an inbox, a wall or an inbox may not read these structures. Similarly, the {\em sender anonymity} of a private message would be broken if the adversary retrieved the public keys $K_i$ or $K_p$ from an inbox. The sender anonymity is preserved as each inbox private message is systematically encrypted with the receiver's public key.

{\em Member unlinkability} also reduces to a confidentiality property. As a storing node does not know the member list it is impossible to link its members.
Only other members of the same two groups would be able to link. 
A determined enough adversary may try to enroll in many groups until she links some principals.
To protect against the later attack a user may create different principal for different groups that he joins.
A user may also renounce unlinkability for some principals that are enrolled in non-critical groups.

Only a very costly attack may break member unlinkability. The attack supposes that the adversary can observe the entire address space at a time (which is equivalent to a central authority). When an administrator just added a joining principal to the group he updates the list structure and sends a message {\tt helo} to the principal's inbox. The adversary may observe these two structures updates occurring at approximately the same time, and thus infer that the principal of the inbox just joined the updated group. Repeating this same attack for a second group would then allow linking the two members. This attack only reveals the principal's inbox address and not its public key. Therefore it does not break member anonymity nor sender anonymity. In addition the attack is extremely costly as it requires to continuously monitor an address space of size $2^{160}$. We therefore consider this attack as unrealistic for our system.

% As these privacy objectives are achieved thanks to confidentiality properties even  adversaries controlling all nodes of the system could not break these properties.

Finally, {\em unlinkability with IP addresses} is achieved by randomly choosing other nodes as proxies as shown in Section~\ref{sec:protocol}. From an adversary node perspective it is thus impossible to decide for a given message if the sender's IP address is the actual address of the sender.

\section{Prototype}
\label{sec:prototype}
%%!TEX TS-program = latex
We have implemented a prototype of our protocol as a proof of concept,
with interfacing capabilities with the Vuze DHT. The goal
of this section is to show that \textit{(i)} our protocol can be
operated on top of a large scale deployed and possibly unmodified
distributed storage infrastructure, and that \textit{(ii)}
performances can be acceptable even in an extreme case of leveraging a
DHT implemented for totally other (best effort) purposes.

\subsection{Settings and challenges}

In order to operate a prototype in a real world setting, we chose to
build on Vuze (previously Azureus), a well known BitTorrent client that
includes a DHT (based on Kademlia) to avoid relying only on
centralized trackers for file distribution. Vuze has been adopted all
around the world, and its DHT is run by around 1.5 millions of users
simultaneously, resulting in various and representative latencies for
a large scale application. It is actually possible to use this DHT for
storing arbitrary data to an also arbitrary address. The first paper
to leverage such an open DHT is describing the Vanish
protocol \cite{usenixsec09geambasu}. Interesting work has been
achieved to parallelize PUT operations on the DHT; as the code of
Vanish experiments is released, we re-use the Vanish interface to the
Vuze DHT. Our prototype uses the latest release of Vuze (4.7.0.0).  Of
course, as we do not control code executed on remote Vuze's nodes, we
can not impose them to implement the verification we presented in
Section \ref{sec:cryptomeans}; they only act as simple nodes
implementing a DHT interface.

\begin{figure}[t!]
\begin{center}
\includegraphics[width=0.9\linewidth]{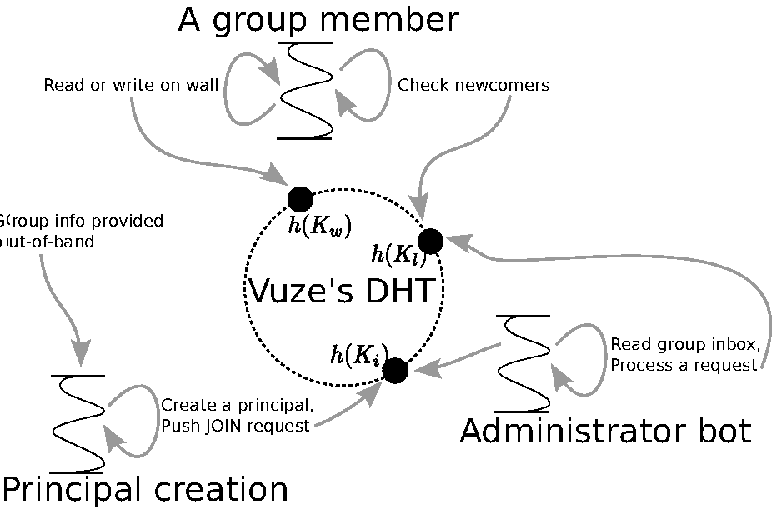}
\end{center}
%\vspace{-0.2cm}
\caption{Scenario considered in our experiments.}
\label{proto}
%\vspace{-0.2cm}
\end{figure}

Prototype is run on a commodity laptop (Intel Core2 Duo at 2.20GHz,
2.0GiB of memory), and using a basic ADSL line (down/up:
18000Kbps/1200Kbps). Code is written in Java and cryptographic
operations use the standard Java Security library.

Using Vuze as a storage back-end for our protocol is challenging,
mostly for two reasons. The first one is that only $512$B of data can
be stored by a PUT ($<key,value>$ insertion). This requires us to
fragment the messages and lists created by our protocol into chunks
to store them, and reversely to re-aggregate those chunks when a GET
operation occurs. The second difficulty is that GET operations are
relatively fast (order of a second), while PUT operations are
prohibitively long (order of
minutes) \cite{usenixsec09geambasu,Falkner:2007:PMU:1298306.1298325}. Concurrency is to be kept in mind
as some operations need to first get a state in the DHT and then write
a result. We chose to operate despite those difficulties, in order to
provide a best effort and worst case illustration of our protocol.

\subsection{Scenario: joining a public group}

Basic protocol functions, described in Section \ref{sec:protocol},
have been implemented; the considered scenario is presented on
Figure \ref{proto}. It consists in creating a group, and to simulate
arrival of join requests to it (following a Poisson process with an
average arrival every $20$ minutes). An administrator bot frequently
retrieves group's inbox in order to always positively process those
requests (this scenario correspond to joining a public group). A group
member is frequently retrieving the list of current group members and
is also reading/writing the group's wall (following another Poisson
process with average $30$ minutes). This scenario has been run
continuously during $2.5$ days ($63$ hours precisely).

\subsection{Protocol evaluation}

\begin{figure}[t!]
\vspace{-1cm}
\begin{center}
\includegraphics[width=0.8\linewidth, angle=-90]{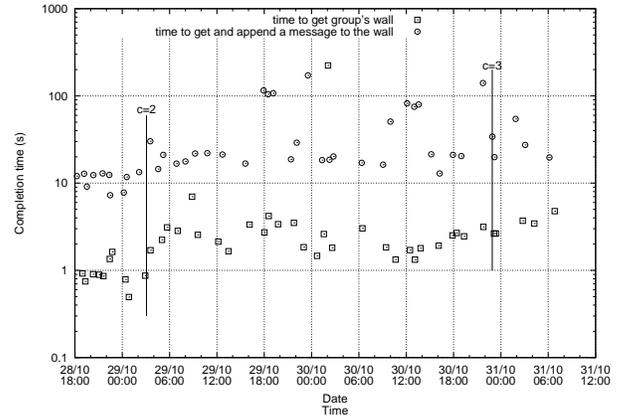}
\end{center}
\vspace{-0.5cm}
\caption{A group member periodically reads and writes the wall, over a period of $2.5$ days. A Poisson process models arrival time of requests, with an average of one operation every $30$ minutes. Operation latencies are reported in seconds, and c indicates the number of chunks composing the wall at a given time.}
\label{wall}
%\vspace{-0.3cm}
\end{figure}

Prior to execute the scenario, we have sequentially created $100$
groups from our laptop, right after a cold start of the Vuze DHT
locally. First $3$ or $4$ creations take a significantly longer time
($2$ or $3$ time) than the average, measured at $16.1$ seconds per
group (standard deviation: $6.2$s). We re-ran the same group creation
process, this time removing operations on the DHT to push data to be
stored; average time drops to $0.66$s per group. This underlines the
fact that network operations totally dominate local structure
manipulations and basic cryptography.

Figure \ref{wall} shows both the time needed by a group member to
retrieves the group's wall, and the time needed to update the wall by
appending few bytes to it (that could correspond to adding a tiny URL
for instance). As the $512$B of allowed storage per insert are quickly
filled by data and integrity information, our message chunking layer
automatically splits and attributes locations in the DHT for the
complete wall to be stored (first chunk still being at $h(K_w)$, while
following ones are stored at $h(K_w.{\tt i})$, with $i$ the $i^{th}$
chunk). Resulting time to read slightly increases, being related to
the time needed by the slowest chunk holder to answer, and thus
finally allowing wall to be reconstructed.  Time needed to modify the
wall is more fluctuating, as contrarily to the read operation (where a
single answer from a chunk replica node is enough), Vuze waits for
replication on the $19$ closest neighbors of the target node to be
complete or to time out. Slow or loaded nodes than slow down the PUT
operation. Please note that we have deliberately chosen to operate in
a worst case setting, as have left the Vuze source code totally
intact, contrarily to paper \cite{usenixsec09geambasu} where some
modifications are made to the Vuze layer itself, making it possible to
decrease storage time from minutes to few seconds.

\begin{figure}[t!]
\vspace{-1cm}
\begin{center}
\includegraphics[width=0.8\linewidth, angle=-90]{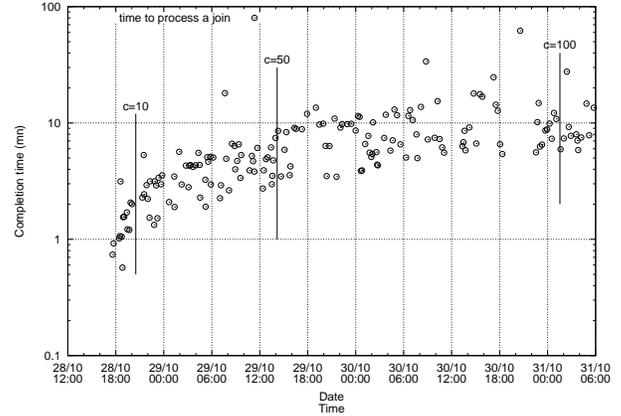}
\end{center}
\vspace{-0.5cm}
\caption{Time needed by the group administrator to process join requests (show in minutes). Requests arrive following a Poisson process with on average one request every $20$ minutes, and are fetched by the administrator as soon as possible.}
\label{joins}
%\vspace{-0.3cm}
\end{figure}

We now have a look at the time needed by the administrator to process
each join request arriving in $h(K_i)$, presented on
Figure \ref{joins}. This constitutes operation of our protocol under
an increasingly unfavorable setting: at the end of this experiment,
$170$ joins have been completed, and the resulting member list is
split into $108$ chunks (even for a total weight of only $54$KB), as
indicated by value c on the figure. This means that when the storage
of structures defined in Section \ref{sec:protocol} for join can be
achieved in a single location, we observe latency in the order of a
minute. Contrariwise, the need to split the structures due to storage
constraints (here the member list) makes our protocol rely on the
slowest set of node chosen to store a chunk; we then reach around $10$
minutes at the end of the run in order to be able to store that list
on the $108$ hosts and replicas. We clearly observe the fact that
operation time for processing join is tied to the number of chunks
constituting the list: while time to write on group's wall
(Figure \ref{wall}) remains mostly steady, join processing time
increases gradually with the number of chuncks (noted c on figure).  A
sub-linear factor increase is nevertheless to be noted,
when considering this number of chunks. Without any dedicated
deployment, simply using Vuze as in, this setting may allow a best
effort and a background group management system to operate, specially
when human interaction is needed to accept or decline requests.

Directions for dedicated deployment and performance improvements
are \textit{(i)} allow a larger storage for $<key,value>$ than the
very restrictive $512$B from Vuze; this would confine performances to
the ones on the very left of those two previous
curves. Secondly \textit{(ii)}, DHT operations should be optimized to
return quickly, as proposed in Vanish implementation; this for
instance includes a quick PUT of an operation result on the
responsible node in the DHT, and then to leave consistency on replica
nodes occur in background.

\section{Related Work}
\label{sec:related}
Socially-enhanced applications are currently the main vectors of the
growth of Internet use.  If means of handling social acquaintances are
developed on an ad-hoc fashion by each new application, to the best
of our knowledge, there is no generic group communication and
management system available in a distributed setting. We believe that
our proposal goes in the direction of genericity, reusability and
applicability. We review main applications that take privacy into
account.

Diaspora \cite{diaspora} proposes a completely distributed approach
for Online Social Networks, in reaction to the recent privacy issues
in Facebook. Today the project is still in alpha-phase and thus only
open to a very restricted number of users. We could not find any
scientific publication on the protocols and security used.

\rev{
OneSwarm~\cite{Isdal:2010:PPD:1851182.1851198} is an extension of
BitTorrent with privacy properties for peer-to-peer data
sharing. OneSwarm builds upon the web of trust, a set of trusted
acquaintances identified and authenticated by individual
public/private key pairs. A user preferably exchanges content and
searches for content within its web of trust and third parties are not
able to participate and monitor a user's web of trust. OneSwarm makes
use of community servers to maintain lists of trusted contacts. In
order to anonymously search and download public content outside its
web of trust, the system supports rewriting source addresses by using
other users of the system as proxies. The system only addresses the
problem of content exchange and thus other types of social-based
application such as collaborative editing or chatting are not
addressed.
}{}

Persona~\cite{Baden:2009:POS:1594977.1592585} proposes the use of
Attribute Based Encryption \cite{10.1109/SP.2007.11} to implement
fine-grained access policies on shared content. According to the set
of groups a user belongs to he can decrypt a given content or
not. Persona supposes that all shared content is encrypted. Thus
revocation leads to reencrypting all contents that were accessible by
the concerned group. The system has no specific requirement on the
storage service that hosts the shared data and thus data may be stored
in a distributed manner. Finally, Persona does not provide any
specific privacy properties such as anonymity or
unlinkability. Instead Persona targets the data
confidentiality within a given group.

Backes et al. \cite{secu_api} present a security api/cryptographic
framework for social applications providing access control on shared
content, privacy of social relations, secrecy of resources, and
anonymity of users. Similar to our approach, users of the system can
create as many pseudonyms as they want and use a different pseudonym
for each relation with the other users of the system. Access control
lists are build upon the created relations. The system uses zero
knowledge protocols to prove the possession of a pseudonym or the
membership of a relation.  Proving a relation membership does 
not reveal the pseudonym. The system has no specific requirement on
the storage service that hosts the shared data and thus data may be
stored in a distributed manner. The proposed system however relies on
a public key infrastructure, which makes it difficult to scale.

Schiavoi et al. \cite{Schiavoni2011} combines gossip-based communication 
protocols and onion routing to build a distributed and private group communication system. 
Groups consist in one or several nodes each knowing the public key of the group. Onion routing is used to achieve sender anonymity, under the hypothesis that each node published a public key. 
Whilst we consider the same threat model regarding privacy properties, we extend the model to a Dolev-Yao adversary \cite{Dolev:1981:SPK:891726} regarding security properties. In addition, we formally validate our security objectives against the latter attacker.
We also address a larger set of operations, such as the creation of a principal and the join mechanism.
Moreover we allow \emph{anonymous} communication from any member of the group to any other member of the same the group.

Finally, one may find the concept of group management related
to \textit{publish-subscribe} mechanisms in distributed
systems~\cite{10385790}. Such systems are typically building multicast
trees among members for message propagation in groups. They differ in
the sense that they are not meant to implement complex and privacy
oriented group management for interaction with social based
applications, but instead focus on simple on-demand multicast.

\section{Conclusion}
\label{sec:conclusion}

In this paper, we have shown the feasibility of distributing group management, in the context of a middleware empowering social-based applications. Previous works have only addressed parts of the distribution process.
We saw that this distribution, as it leverages resources scattered among many authorities, constrains achievable security objectives. Yet, this paper shows that reasonable security and privacy properties can be reached. Our system also removes the control and lock of a single operator or organization on the group dynamic, improving state of the art in the direction of scalable and reusable application.

Future work will focus on the key renewal protocol to support cases where the group administrator does not know the group members. Development or adaptation of a load balancing
mechanism for handling popular groups is also to be achieved.
Finally, we envision validating the privacy properties of our protocol
formally using recent extensions to Scyther \cite{Scyther}.

\bibliographystyle{abbrv}
\bibliography{SecureGroupManagement}

\clearpage{}
\onecolumn

\appendix[HLPSL code]

We provide here parts of the HLPSL code used for checking security properties.
The first portion of code is used for checking weak authentication of the entity updating a captured node.

{\scriptsize
\begin{verbatim}
% Message types are coded by naturals, 10:root 
% 11 is not a real protocol message, it is just here to ease the simulation

role cre    (CRE,DHT:agent, Hash:hash_func, SND,RCV:channel(dy)) played_by CRE def=
local       State:nat, Kr:public_key
init        State:= 0
transition 
cr.         State = 0 /\ RCV(start) =|> State':= 1 /\ Kr':=new() /\
            SND(Hash(Kr').{10.0.Kr'}_inv(Kr')) 
rc.         State = 1 /\ RCV(11) =|> State':= 2
ur.         State = 2 =|> State':= 3 /\ SND(Hash(Kr).{10.1.Kr}_inv(Kr)) /\ 
            witness(DHT,CRE,dht_cre_kr,Kr)
end role

role dht    (CRE,DHT:agent, Hash:hash_func, SND,RCV:channel(dy)) played_by DHT def=
local       State:nat, Kr:public_key
init        State:= 0
transition 
cr.         State = 0 /\ RCV(Hash(Kr').{10.0.Kr'}_inv(Kr')) =|> State':= 1 /\ SND(11)
ur.         State = 1 /\ RCV(Hash(Kr).{10.1.Kr}_inv(Kr)) =|> State':= 2 /\
            wrequest(DHT,CRE,dht_cre_kr,Kr)
end role

role session(CRE,DHT:agent, Hash:hash_func) def=
local       SND,RCV:channel(dy)
composition 
            cre(CRE,DHT,Hash,SND,RCV) /\ dht(CRE,DHT,Hash,SND,RCV)
end role

role environment() def=
const       g,d:agent, h:hash_func, dht_cre_kr:protocol_id
intruder_knowledge={g,d,h}
composition
            session(g,d,h)
end role

goal weak_authentication_on _cre_kr end goal

environment()
\end{verbatim}
}

\begin{figure}[h]
\begin{center}
\includegraphics[width=0.9\linewidth]{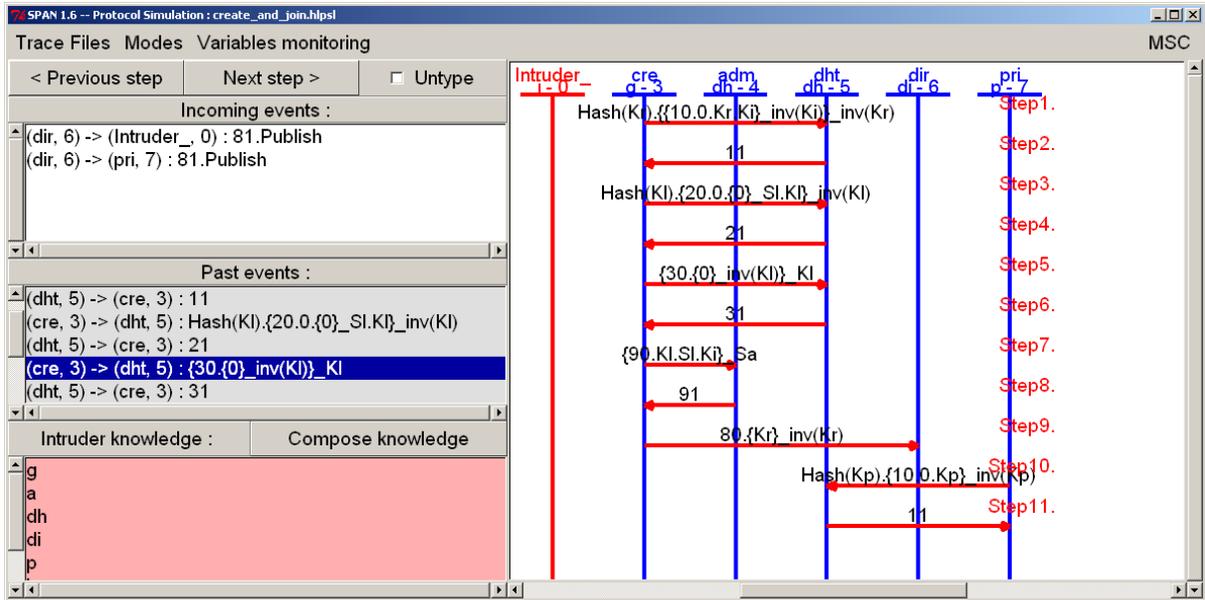}
\end{center}
\caption{Simulation of the second portion of code, within the AVISPA + SPAN framework.}
\label{fig:span}
\end{figure}

The second portion of code is used for proving secrecy properties, in particular the secrecy of the key $K_l$ for all but the administrator and the creator.  
The figure \ref{fig:span} corresponds to a simulation of the second portion of code for creating a group and a principal and then forging a join request.

{\scriptsize 
\begin{verbatim}
% Message types are coded by naturals
% 10:root 20:list 30:once 40:wall 50:join 60:helo 70:mess 80:name (for directory) 90:admin

role cre    (CRE,ADM,DHT1,CHT2:agent, Sa:symmetric_key, Hash:hash_func, SND,RCV:channel(dy)) played_by CRE def=
local       State,Cr,Cl,Cw:nat, Kr,Ki,Kl,Kw:public_key, Sl,Sw:symmetric_key
init        State:= 0
transition 
%           Capture Root
cr          State = 0 /\ RCV(start) =|> State':= 1 /\ Kr':=new() /\ Ki':=new() /\
            SND(Hash(Kr').{{10.0.Kr'.Ki'}_inv(Ki')}_inv(Kr'))
rc.         State = 1 /\ RCV(11) =|> State':= 2
%           Capture List
cl.         State = 2 =|> State':=3 /\ Kl':=new() /\ Sl':=new() /\
            secret(Kl',secr_kl,{CRE,ADM}) /\
            SND(Hash(Kl').{20.0.{0}_Sl'.Kl'}_inv(Kl'))
lc.         State = 3 /\ RCV(21) =|> State':= 4
%           Set Once
so.         State = 4 =|> State':=5 /\ SND({30.{0}_inv(Kl)}_Kl)
os.         State = 5 /\ RCV(31) =|> State':=6
%           Securely send Kl Sl Ki to an admin (assuming a dedicated channel Sa)
sa.         State = 6 =|> State':=7 /\ SND({90.Kl.Sl.Ki}_Sa)
as.         State = 7 /\ RCV(91) =|> State':=8
%           Publish in Directory
pd.         State = 8 =|> State':=9 /\ SND(80.{Kr}_inv(Kr))
end role

role dir    (CRE,DIR,PRI:agent, SND,RCV:channel(dy)) played_by DIR def=
local       State:nat, Publish:message
init        State:=0
transition
%           Recieve and Publish
rp.         State = 0 /\ RCV(80.Publish') =|> State':=0 /\ SND(81.Publish')
end role

% this part of the *is* under the control of the adversary
role dht1   (CRE,DHT1,DHT2:agent, Hash:hash_func, SND,RCV:channel(dy)) played_by DHT1 def=
local       State,Cr,Cl,Cw:nat, Kr,Ki,Kl,Kw,Kj:public_key, Sl,Sw:symmetric_key, Mi:message
init        State:= 0
transition 
%           Receive Once
cl.         State = 0 /\ RCV({30.{0}_inv(Kl)}_Kl) =|> State':= 0 /\ SND(31)
%           Receive any message in inbox
ri.         State = 0 /\ RCV(70.Mi') =|> State':=0 /\ SND(70.Mi')
end role

% this part of the dht *is not* under the control of the adversary
role dht2   (CRE,DHT1,DHT2:agent, Hash:hash_func, SND,RCV:channel(dy)) played_by DHT2 def=
local       State,Cr,Cl,Cw:nat, Kr,Ki,Kl,Kw,Kj:public_key, Sl,Sw:symmetric_key
init        State:= 0
transition 
%           Capture Root (group)
cr.         State = 0 /\ RCV(Hash(Kr').{{10.0.Kr'.Ki'}_inv(Ki')}_inv(Kr')) =|> State':= 0 /\ SND(11)
pr.         State = 0 /\ SND(Hash(Kr).{{10.0.Kr.Ki}_inv(Ki)}_inv(Kr)) =|> State':= 0 /\ SND(82)
%           Capture Root (principal)
cr.         State = 0 /\ RCV(Hash(Kr').{10.0.Kr'}_inv(Kr')) =|> State':= 0 /\ SND(11)
%           Update Root
ur.         State = 0 /\ RCV(Hash(Kr).{{10.1.Kr.Ki}_inv(Ki)}_inv(Kr)) =|> State':= 1 /\ SND(12)
%           Capture List
cl.         State = 0 /\ RCV(Hash(Kl').{20.0.{0}_Sl'.Kl'}_inv(Kl')) =|> State':= 0 /\ SND(21)
end role


role adm    (CRE,ADM,DHT1,DHT2:agent, Sa:symmetric_key, Hash:hash_func, SND,RCV:channel(dy)) played_by ADM def=
local       State,Cr,Cl,Cw:nat, Kr,Ki,Kl,Kw,Kj:public_key, Sl,Sw:symmetric_key
init        State:= 0
transition 
%           Becomming an admin by recieving Kl Sl Ki
cr.         State = 0 /\ RCV({90.Kl'.Sl'.Ki'}_Sa) =|> State':= 1 /\ SND(91)
end role


role pri    (PRI,DIR,DHT1,DHT2:agent, Hash:hash_func, SND,RCV:channel(dy)) played_by PRI def=
local       State,Cr:nat, Kp,Kj,Kr,Ki:public_key, Sl,Sw:symmetric_key, Mi:message
init        State:= 0
transition 
%           Create Principal
cp.         State = 0 /\ RCV(start) =|> State':=1 /\ Kp':=new() /\ SND(Hash(Kp').{10.0.Kp'}_inv(Kp'))
pc.         State = 1 /\ RCV(11) =|> State':=2
%           Retrieve group info from the directory
rd.         State = 2 /\ RCV(81.{Kr'}_inv(Kr')) =|> State':=3
%           Retrieve group inbox from the group root
ri.         State = 3 /\ RCV(82) =|> RCV({{10.0.Kr'.Ki'}_inv(Ki')}_inv(Kr')) /\ State':=4 
%           Request Join
%rj.        State = 4 /\ Kj':=new() /\ SND({{{60.Kp.Kj
end role


role session(CRE,ADM,DHT1,DHT2,DIR,PRI:agent, Sa:symmetric_key, Hash:hash_func) def=
local       SND,RCV:channel(dy)
composition 
            cre(CRE,ADM,DHT1,DHT2,Sa,Hash,SND,RCV) /\ 
            adm(CRE,ADM,DHT1,DHT2,Sa,Hash,SND,RCV) /\
            dht1(CRE,DHT1,DHT2,Hash,SND,RCV)       /\ 
            dht2(CRE,DHT1,DHT2,Hash,SND,RCV)       /\ 
            dir(CRE,DIR,PRI,SND,RCV)               /\
            pri(PRI,DIR,DHT1,DHT2,Hash,SND,RCV) 
end role

role environment() def=
const       g,a,d1,d2,di,p:agent, sa:symmetric_key, h:hash_func, secr_kl:protocol_id
intruder_knowledge={g,a,d1,d2,di,p,h}
composition
            session(g,a,d1,d2,di,p,sa,h)
end role

goal secrecy_of secr_kl end goal

environment()

\end{verbatim}
}

\end{document}